\documentclass[12pt]{article}

\usepackage[authoryear]{natbib}
\usepackage{amssymb,amsthm,epsfig}
\usepackage[centertags]{amsmath}
\usepackage{bm}
\usepackage{lineno}

\newcommand{\stackunder}[2]{ \renewcommand{\arraystretch}{0.4}
\displaystyle \begin{array}[t]{c}  {#1}\\_{#2}\end{array}
\renewcommand{\arraystretch}{1}}
\newcommand{\vvec}{\textrm{vec}}
\newcommand{\tr}{\textrm{tr}}
\newcommand{\Es}{\mathbb{E}}

\newcommand{\sumdir}{ \hspace{-0.1cm} \stackrel{n}{\stackunder{\oplus}{i=1}} \hspace{-0.3cm}}

\newtheorem*{teorema}{{\sc Theorem}}

\title{Bias correction in a multivariate normal regression model with general parameterization} 
\author{Alexandre G.~Patriota,\quad Artur J.~Lemonte\\  
{\small {\em Departamento de Estat\'istica, Universidade de S\~ao Paulo, Rua do Mat\~ao, 1010,}}\\
{\small {\em S\~ao Paulo/SP, 05508-090, Brazil}}}
\date{}

\begin{document}
\maketitle

\begin{abstract}
This paper develops a bias correction scheme for a multivariate normal model under a 
general parameterization. In the model, the mean vector and the covariance 
matrix share the same parameters. It includes many important regression
models available in the literature as special cases, such as (non)linear regression, errors-in-variables 
models, and so forth. Moreover, heteroscedastic situations may also be studied
within our framework. We derive a general expression for the second-order biases of maximum likelihood 
estimates of the model parameters and show that it is always possible to obtain the second order bias
by means of ordinary weighted lest-squares regressions. We enlighten such general expression
with an errors-in-variables model and also conduct some simulations 
in order to verify the performance of the corrected estimates. The simulation
results show that the bias correction scheme yields nearly unbiased estimators.
We also present an empirical ilustration.\\ 

\end{abstract}

\section{Introduction}

Applications of multivariate normal models are commonly found in the literature. There are
simple models that do not require asymptotic approximations for the maximum likelihood 
estimators. Nevertheless, in the majority of problems, the estimation procedure in such 
multivariate normal models embrace on the asymptotic theory. For instance, it is hard to 
compute the exact distribution of maximum likelihood estimators (MLEs) for nonlinear 
multivariate regressions, errors-in-variables models and many others when the sample size is finite (practical situations).
Then, in the practical applications, the asymptotic distribution of the MLE is used as an approximation 
to its exact distribution. It considerable simplifies the inferential process. In general, under some 
regularity conditions, the MLEs are consistent and efficient, i.e., asymptotically, their biases converge 
to zero and their variance-covariance matrices approach the
inverse of the Fisher information. Moreover, under such regularity 
conditions, the MLEs are asymptotically normally distributed. Although the MLEs have these important 
features, they may be strongly biased for small or even moderate samples sizes when more 
complex models are considered. Thus, a bias correction can play an important role to improve 
the estimation of the model parameters.

An important area of research in statistics is the study of the finite-sample behavior
of MLEs. It is well konwn that MLEs are oftentimes
biased, thus displaying systematic error. This is not a serious problem
for relatively large sample sizes, since the bias is typically of order $\mathcal{O}(n^{-1})$, whereas
the asymptotic standard errors are of order $\mathcal{O}(n^{-1/2})$. However, for small or
even moderate values of the sample size $n$, bias can constitute a problem. Thus,
availability of formulae for its approximate computation is important for accurate estimation
of many models that are used in a number of applications.
Bias correction of MLEs is particularly important when the sample size, or the
total information, is small \citep{VascCribari2005}.

Bias adjustment has been extensively studied in the statistical literature.
\cite{Box1971} gives a general expression for the $n^{-1}$ bias in multivariate nonlinear models
where covariance matrices are known. For nonlinear regression models, \cite{CTWei(1986)}
relate bias to the position of the explanatory variables in
the sample space. \cite{CordMcC(1991)} give general matrix formulae
for bias correction in generalized linear
models. \cite{CorVasc1997}
obtained general matrix formulae for bias correction in multivariate nonlinear regression models with
normal errors, while \cite{Vasccor1997}
obtained general formulae for bias in multivariate nonlinear heteroscedastic regression.
Also, \cite{CordVasc(1999)} obtained second order biases of the maximum likelihood estimators
in von Mises regression models, while \cite{CordFUVasc2000}
obtained bias correction for symmetric nonlinear regression models.
\cite{VascCor2000} obtained bias correction
for multivariate nonlinear Student $t$ regression models. 
More recently, \cite{VascCribari2005} obtained bias correction
in a new class of beta regressions. \cite{CorDem2008}
derive formulae for the second-order biases of the quasi-maximum likelihood
estimators, while \cite{CorToy(2008)} derive general formulae for the second-order biases
in generalized nonlinear models with dispersion covariates.
 
In this paper we study a multivariate normal model with general parameterization and derive
the second-order biases of the maximum likelihood estimates. Here, the general parameterization 
means a sort of unification of several important models which can be constructed using the 
multivariate normal model. For instance, the multivariate nonlinear regressions
studied by \cite{CorVasc1997} and their heteroscedastic version \citep{Vasccor1997}
are just particular cases of our proposal. In this paper we propose a model in which the mean
$\bm{\mu}$ and the variance $\bm{\Sigma}$ of the observed variables are indexed by the same vector of 
parameters, say $\bm{\theta}$. The existing works on bias correction assume that the mean and variance
do not share any parameters, however, in errors-in-variables models, for example, 
this assumption is not realistic. Indeed, that assumption makes the computation of the bias formulae
less complicated, but it restricts the applicability of the approach to a special class of models.
In view of that, the main goal of this article is to extend the bias correction to a wide class 
of multivariate models which has not yet been considered in the statistical literature.

The outline of the paper is as follows. Section 2 presents the main model and computes the first
three derivatives of the log-likelihood function and their expectations. In Section 3, we present 
matrix formulae for the second-order biases of the MLEs for the general model.
In Section 4, we present some useful examples of the proposed formulation.
Monte Carlo simulation results are presented and
discussed in Section~\ref{simulation}. The numerical results show that the bias correction we
derive is effective in small samples; it delivers estimators that are nearly unbiased
and display superior finite-sample behavior. Section~\ref{application}
contains an empirical ilustration. Finally, Section~\ref{conclusion} concludes the
paper.

\section{Model specification}\label{model-specification}

We consider the situation in which $n$ independent multivariate random variables $\bm{Y}_{1},
\ldots, \bm{Y}_{n}$ are observed and the number of responses measured in each observation
is $q$. We also admit that auxiliary covariates can be observed, say $\bm{X}_1, \ldots, \bm{X}_n$.
 The multivariate regression model can then be represented as
\begin{equation}\label{MainModel}
\bm{Y}_{i} = \bm{\mu}_{i}(\bm{\theta}) + \bm{u}_{i},\quad i=1,2,\ldots,n,
\end{equation}
where $\bm{Y}_{i}$ is a $q\times 1$ vector of dependent variables, $\bm{\mu}_{i}(\bm{\theta})\equiv
\bm{\mu}_{i}(\bm{\theta},\bm{X}_{i})$ is a mean function (the shape is assumed known) which is 
 three times continuously differentiable with respect to each element of $\bm{\theta}$ and 
$\bm{X}_{i}$ is an $m\times 1$ vector of known explanatory variables associated with the $i^{th}$ 
observed response $\bm{Y}_{i}$.
Also, $\bm{\theta}$ is a $p\times 1$ vector of unknown parameters of interest. Additionally, as the
foundation for estimation by maximum likelihood and hypothesis testing, we assume
that the independent random errors $\bm{u}_{i}$'s follow a multivariate
normal $\mathcal{N}_{q}(\bm{0}, \bm{\Sigma}_i(\bm{\theta}))$ distribution, where
$\bm{\Sigma}_i(\bm{\theta})$ is a $q\times q$ nonsingular covariance matrix and the
entries of $\bm{\Sigma}_i(\bm{\theta})$ are assumed three times continuously differentiable
in each element of $\bm{\theta}$. We are assuming, in addition, that the functions $\bm{\mu}_i(\bm{\theta})$
and $\bm{\Sigma}_i(\bm{\theta})$ are defined in a convenient way since $\bm{\theta}$ must be identifiable in model~(\ref{MainModel}). 

The class of models presented above is very rich and includes many important
regression models. For example, in an errors-in-variables model we observe two variables,
namely $Y_i$ and $X_i$ whose relationship is given by
\begin{equation}\label{EVM}
Y_i = \alpha + \beta x_i + e_i\quad{\rm and}\quad X_i = x_i + u_i,
\end{equation}
where $x_i \sim \mathcal{N}(\mu_x, \sigma_x^2)$, $e_i \sim \mathcal{N}(0,\sigma^2)$ 
and $u_i \sim \mathcal{N}(0, \sigma_u^2)$, with $\sigma_u^2$ known and, additionally,
$x_i$, $e_i$ and $u_i$  are mutually uncorrelated. Then, denoting $\bm{Y}_i = (Y_i, X_i)^{\top}$
and $\bm{\theta} = (\alpha, \beta, \mu_x, \sigma_x^2, \sigma^2)^{\top}$ we have that 
$\bm{Y}_i \sim \mathcal{N}_2(\bm{\mu}(\bm{\theta}), \bm{\Sigma}(\bm{\theta}))$, where
\[
\bm{\mu}(\bm{\theta}) = 
\begin{pmatrix}
\alpha + \beta\mu_{x} \\
\mu_{x}
\end{pmatrix} \quad \mbox{and} \quad
\bm{\Sigma}(\bm{\theta})=
\begin{pmatrix}
\beta^2 \sigma_x^2 + \sigma^2 & \beta \sigma_x^2\\
\beta \sigma_x^2 & \sigma_x^2 + \sigma_u^2
\end{pmatrix}.
\]
This is a simple linear regression in which the covariate is subject to measurement errors.
This is a good example where the usual approach 
(assuming that $\bm{\Sigma}$ and $\bm{\mu}$ do not share any parameter) is not applicable.
Measurement error models have been largely used in epidemiology \citep{Kulathinal2002,deCastro2008}, astrophysics 
\citep{Akritas1996, Kelly2007, Kellyetal2008} and analytical chemistry
\citep{ChengRiu2006} to avoid inconsistent 
estimators \citep[see][for further details]{Fuller}.  Other special 
cases of model~(\ref{MainModel}) are: 
multivariate heteroscedastic nonlinear errors-in-variables models, multivariate nonlinear heteroscedastic 
models, univariate nonlinear models, factor analysis, mixed models and so on. As can be seen, model 
(\ref{MainModel}) can encompass a wide class of models.

To simplify the notation, define
$\bm{Y}= \vvec(\bm{Y}_1,\bm{Y}_{2},\ldots,\bm{Y}_n)$,
$\bm{\mu} = \vvec(\bm{\mu}_1(\bm{\theta}),\ldots,\bm{\mu}_n(\bm{\theta}))$,
$\bm{\Sigma} = {\rm diag}\{\bm{\Sigma}_1(\bm{\theta}),\ldots,\bm{\Sigma}_n(\bm{\theta})\}$
and $\bm{u} = \bm{Y} - \bm{\mu}$, where vec$(\cdot)$ is the vec operator, which transforms a matrix into a 
vector by stacking the columns of the matrix one underneath the other.
Then, the log-likelihood associated with (\ref{MainModel}),
apart from an unimportant constant, is 
\begin{equation}\label{log-likelihood}
\ell(\bm{\theta})\propto -\dfrac{1}{2} \log{|\bm{\Sigma}|}
-\dfrac{1}{2}\tr\{\bm{\Sigma}^{-1}\bm{u}\bm{u}^{\top}\}.
\end{equation}
We make some assumptions \citep[][Ch.~9]{CoxHinkley1974} on the behavior of
$\ell(\bm{\theta})$ as the sample size $n$ approaches infinity, such as the
regularity of the first three derivatives of $\ell(\bm{\theta})$ with respect to $\bm{\theta}$
and the uniqueness of the MLE of $\bm{\theta}$, $\widehat{\bm{\theta}}$. We now introduce the 
following total log-likelihood derivatives, in which the 
indices $r$, $s$ and $t$ range from $1$ to $p$. Let $U_{r} = \partial\ell(\bm{\theta})/\partial\theta_{r}$,
$U_{sr} = \partial^{2}\ell(\bm{\theta})/\partial\theta_{s}\partial\theta_{r}$ and 
$U_{tsr} = \partial^{3}\ell(\bm{\theta})/\partial\theta_{t}\partial\theta_{s}\partial \theta_{r}$ be the first
three derivatives of $\ell(\bm{\theta})$. The standard notation for the moments of those 
log-likelihood derivatives is used \citep{Lawley1956}, namely: $\kappa_{sr} = \Es(U_{sr})$, 
$\kappa_{s,r} = \Es(U_{s}U_{r})$, $\kappa_{tsr} = \Es(U_{tsr})$
and so on. Furthermore, we define the derivative of $\kappa_{sr}$ with respect to $\theta_{t}$ as 
$\kappa_{sr}^{(t)} = \partial\kappa_{sr}/\partial\theta_{t}$.
Not all $\kappa$'s are functionally independent; e.g., 
$\kappa_{s,r} = -\kappa_{sr}$, which is the typical element of the information matrix 
$\bm{K}_{\bm{\theta}}$, assumed to be nonsingular. All $\kappa$'s
refer to a total over the sample and are, in general, of order $n$. Finally,
let $\kappa^{s,r}$ denote the corresponding element of $\bm{K}_{\bm{\theta}}^{-1}$.

Define the following quantities:
\[
\bm{a}_r = \frac{\partial \bm{\mu}}{\partial \theta_{r}}, \quad
\bm{a}_{sr} = \frac{\partial^2 \bm{\mu}}{\partial \theta_{s} \partial\theta_{r}}, \quad
\bm{a}_{tsr} = \frac{\partial^3 \bm{\mu}}{\partial \theta_{t}\partial\theta_{s} \partial\theta_{r}}, \quad
\bm{C}_{r} = \frac{\partial \bm{\Sigma}}{\partial \theta_{r}}, \quad
\bm{C}_{sr} = \frac{\partial^2 \bm{\Sigma}}{\partial \theta_{s}\partial\theta_{r}},
\]
\[
\bm{C}_{tsr} = \frac{\partial^3\bm{\Sigma}}{\partial\theta_{t}\partial\theta_{s}\partial\theta_{r}},
\quad\bm{A}_{r} = \frac{\partial\bm{\Sigma}^{-1}}{\partial\theta_{r}} = -\bm{\Sigma}^{-1} \bm{C}_r\bm{\Sigma}^{-1}
\quad{\rm and}\quad
\bm{A}_{sr} = \frac{\partial\bm{A}_{r}}{\partial\theta_{s}},
\]
where $r,s,t = 1,2,\ldots,p$. We assume that these derivatives do exist. To compute the 
derivatives of $\ell(\bm{\theta})$ we make use of methods in matrix differential calculus, as describe in
\cite{MagnusNeudecker}. Thus, the first derivative of (\ref{log-likelihood}) with respect to the $r^{th}$
element of $\bm{\theta}$ is 
\[
U_{r} = \dfrac{1}{2} \tr\{\bm{A}_{r}(\bm{\Sigma} - \bm{u}\bm{u}^{\top})\} +
\tr\{\bm{\Sigma}^{-1}\bm{a}_{r} \bm{u}^{\top}\}.
\]
By using some simple matrix properties, the score function for $\bm{\theta}$ can be written in matrix form as
\[
\bm{U}_{\bm{\theta}}\equiv\bm{U}_{\bm{\theta}}(\bm{\theta}) = \widetilde{\bm{D}}^{\top}\bm{\Sigma}^{-1}\bm{u}
-\frac{1}{2}\widetilde{\bm{V}}^{\top}\widetilde{\bm{\Sigma}}^{-1}\vvec(\bm{\Sigma} - \bm{u}\bm{u}^{\top}), 
\]
where $\widetilde{\bm{D}} = (\bm{a}_1, \ldots, \bm{a}_{p})$, $\widetilde{\bm{V}}
= (\vvec(\bm{C}_{1}),\ldots, \vvec(\bm{C}_p))$, $\widetilde{\bm{\Sigma}} = \bm{\Sigma} \otimes \bm{\Sigma}$
and $\otimes$ is the Kronecker product. Let
\begin{equation}\label{matrix-score}
\widetilde{\bm{F}} = 
\begin{pmatrix}
\widetilde{\bm{D}}\\
\widetilde{\bm{V}}
\end{pmatrix},\quad
\widetilde{\bm{H}} =
\begin{pmatrix}
\bm{\Sigma} & \bm{0}\\
\bm{0} & 2\widetilde{\bm{\Sigma}}
\end{pmatrix}^{-1}
\quad{\rm and}\quad
\widetilde{\bm{u}} =
\begin{pmatrix}
\bm{u}\\
-\vvec(\bm{\Sigma} - \bm{u}\bm{u}^{\top})
\end{pmatrix}.
\end{equation}
Then, note that the score function can be written as
\[
\bm{U}_{\bm{\theta}} = \widetilde{\bm{F}}^{\top}\widetilde{\bm{H}}\widetilde{\bm{u}}.
\]
The second and third derivatives are given, respectively, by
\begin{align*}
U_{sr} &= \frac{1}{2}\tr\{(\bm{A}_{s}\bm{\Sigma}\bm{A}_{r} +
\bm{A}_{r}\bm{\Sigma}\bm{A}_{s}-\bm{\Sigma}^{-1}\bm{C}_{sr}\bm{\Sigma}^{-1})(\bm{\Sigma} - \bm{u}\bm{u}^{\top})\}\\
&\quad+\frac{1}{2}\tr\{\bm{A}_{r}(\bm{C}_{s} + \bm{a}_{s}\bm{u}^{\top} + \bm{u}\bm{a}_{s}^{\top})\}
+\tr\{(\bm{A}_{s}\bm{a}_{r} + \bm{\Sigma}^{-1}\bm{a}_{sr})\bm{u}^{\top} - \bm{\Sigma}^{-1}\bm{a}_{r}\bm{a}_{s}^{\top}\}
\end{align*}
and
\[
U_{tsr} =  U_{tsr}^{(1)} +U_{tsr}^{(2)} +U_{tsr}^{(3)} +U_{tsr}^{(4)},
\]
where
\begin{align*}
U_{tsr}^{(1)} &= -\frac{1}{2}\tr\{(\bm{\Sigma}^{-1}\bm{C}_{tsr}\bm{\Sigma}^{-1} +
\bm{\Sigma}^{-1}\bm{C}_{rs}\bm{A}_{t} + \bm{A}_{t}\bm{C}_{sr}\bm{\Sigma}^{-1})(\bm{\Sigma} - \bm{u}\bm{u}^{\top})\\
&\quad +\bm{\Sigma}^{-1}\bm{C}_{sr}\bm{\Sigma}^{-1}\bm{C}_{t}
+\bm{\Sigma}^{-1}\bm{C}_{sr}\bm{\Sigma}^{-1}(\bm{a}_{t}\bm{u}^{\top} + \bm{u}\bm{a}_{t}^{\top})\},
\end{align*}
\begin{align*}
U_{tsr}^{(2)} &= \frac{1}{2}\tr\{(\bm{A}_{st}\bm{\Sigma}\bm{A}_{r}
+\bm{A}_{s}(\bm{C}_{t}\bm{A}_{r} + \bm{\Sigma}\bm{A}_{rt}))(\bm{\Sigma} - \bm{u}\bm{u}^{\top})\\
&\quad +(\bm{A}_{rt}\bm{\Sigma}\bm{A}_{s} 
+\bm{A}_{r}(\bm{C}_{t}\bm{A}_{s} + \bm{\Sigma}\bm{A}_{st}))(\bm{\Sigma} - \bm{u}\bm{u}^{\top})\\
&\quad +(\bm{A}_{s}\bm{\Sigma}\bm{A}_{r} + \bm{A}_{r}\bm{\Sigma}\bm{A}_{s}) 
(\bm{C}_{t} + \bm{a}_{t}\bm{u}^{\top} + \bm{u}\bm{a}_{t}^{\top})\},
\end{align*}
\[
U_{tsr}^{(3)} = \frac{1}{2}\tr\{\bm{A}_{rt}(\bm{C}_{s} + \bm{a}_{s}\bm{u}^{\top} + \bm{u}\bm{a}_{s}^{\top})
+\bm{A}_{r}(\bm{C}_{ts} + \bm{a}_{ts}\bm{u}^{\top} - \bm{a}_{s}\bm{a}_{t}^{\top}
+ \bm{a}_{t}\bm{a}_{s}^{\top} + \bm{u}\bm{a}_{ts}^{\top})\}
\]
and
\begin{align*}
U_{tsr}^{(4)} &= \tr\{(\bm{A}_{st}\bm{a}_{r} + \bm{A}_{s}\bm{a}_{rt} + \bm{A}_{t}\bm{a}_{sr}
+ \bm{\Sigma}^{-1}\bm{a}_{srt})\bm{u}^{\top}-(\bm{A}_{s}\bm{a}_{r}+ \bm{\Sigma}^{-1}\bm{a}_{sr})\bm{a}_{t}^{\top}
\}\\
&\quad - \tr\{\bm{\Sigma}^{-1}\bm{a}_{r}\bm{a}_{ts}^{\top} + \bm{A}_{t}\bm{a}_{r}\bm{a}_{s}^{\top}
+ \bm{\Sigma}^{-1}\bm{a}_{s}\bm{a}_{tr}^{\top}\}.
\end{align*}
Note that $\Es(\bm{u}) = \bm{0}$ and $\Es(\bm{u}\bm{u}^{\top})= \bm{\Sigma}$. Knowing these properties, the
expectation of $U_r$, $U_{sr}$ and $U_{tsr}$ are easily obtained. 
The quantities $\kappa_{sr}$, $\kappa_{tsr}$ and $\kappa_{ts}^{(r)}$
($r,s,t = 1,2,\ldots,p$) are given, respectively, by
\begin{equation}\label{krs}
\kappa_{sr} = \frac{1}{2}\tr\{\bm{A}_{r}\bm{C}_{s}\} -  \bm{a}_{s}^{\top}\bm{\Sigma}^{-1}\bm{a}_{r},
\end{equation}
\begin{align}\label{krst}
\begin{split}
\kappa_{tsr} &= \tr\{(\bm{A}_r \bm{\Sigma}\bm{A}_s + \bm{A}_s\bm{\Sigma}\bm{A}_r) \bm{C}_t\}
+ \frac{1}{2} \tr\{ 
 \bm{A}_s\bm{C}_{tr} + \bm{A}_r\bm{C}_{ts}+\bm{A}_t\bm{C}_{sr} \}\\
&- (\bm{a}_t^{\top} \bm{A}_s \bm{a}_r + \bm{a}_s^{\top}\bm{A}_t \bm{a}_r +\bm{a}_s ^{\top}\bm{A}_r \bm{a}_t+
\bm{a}_t^{\top}\bm{\Sigma}^{-1}\bm{a}_{sr} 
+\bm{a}_{ts}^{\top} \bm{\Sigma}^{-1}\bm{a}_{r} +
\bm{a}_s^{\top}\bm{\Sigma}^{-1}\bm{a}_{tr})
\end{split}
\end{align}
and 
\begin{align}\label{krs(t)}
\begin{split}
\kappa_{ts}^{(r)} &= \frac{1}{2}\tr\{(\bm{A}_{r}\bm{\Sigma}\bm{A}_{s} +
\bm{A}_{s}\bm{\Sigma}\bm{A}_{r})\bm{C}_t + \bm{A}_t\bm{C}_{rs}+\bm{A}_{s}\bm{C}_{rt}\}\\
&\quad  - ( \bm{a}_{rt}^{\top}\bm{\Sigma}^{-1}\bm{a}_{s}
+\bm{a}_{t}^{\top}\bm{A}_{r}\bm{a}_{s} + \bm{a}_{t}^{\top}\bm{\Sigma}^{-1}\bm{a}_{rs}).
\end{split}
\end{align}

Again, by using some matrix properties on expression~(\ref{krs}), we can
written the expected Fisher information as
\[
\bm{K}_{\bm{\theta}}\equiv\bm{K}_{\bm{\theta}}(\bm{\theta}) = \widetilde{\bm{D}}^{\top}\bm{\Sigma}^{-1} \widetilde{\bm{D}}
+\frac{1}{2}\widetilde{\bm{V}}^{\top}\widetilde{\bm{\Sigma}}^{-1}\widetilde{\bm{V}}.
\]
Using $\widetilde{\bm{F}}$ and $\widetilde{\bm{H}}$ given in~(\ref{matrix-score}), we can write
$\bm{K}_{\bm{\theta}}$ in the form
\begin{equation}\label{InfFisher}
\bm{K}_{\bm{\theta}} = \widetilde{\bm{F}}^{\top}\widetilde{\bm{H}}\widetilde{\bm{F}}.
\end{equation}

The MLE $\widehat{\bm{\theta}}$ satisfy the equation $\bm{U}_{\bm{\theta}} = \bm{0}$. 
After some matrix manipulations, the Fisher scoring method can be used to estimate $\bm{\theta}$ by
iteratively solving the equation
\begin{equation}\label{Fisher-Scoring}      
(\widetilde{\bm{F}}^{(m)\top}\widetilde{\bm{H}}^{(m)}\widetilde{\bm{F}}^{(m)})\bm{\theta}^{(m+1)} = 
\widetilde{\bm{F}}^{(m)\top}\widetilde{\bm{H}}^{(m)}\widetilde{\bm{u}}^{*(m)},\quad
m = 0, 1, 2,\ldots,
\end{equation}
where $\widetilde{\bm{u}}^{*(m)} = \widetilde{\bm{F}}^{(m)}\bm{\theta}^{(m)}
+ \widetilde{\bm{u}}^{(m)}$. Each loop, through the iterative scheme~(\ref{Fisher-Scoring}),
consists of an iterative re-weighted least squares algorithm to optimize the log-likelihood~(\ref{log-likelihood}).
Using equation~(\ref{Fisher-Scoring}) and any software ({\tt MAPLE}, {\tt MATLAB}, {\tt Ox}, {\tt R}, {\tt SAS})
with a weighted linear regression routine one can compute the MLE $\widehat{\bm{\theta}}$ iteratively.
It is also noteworthy that the MLE in even much complex models, such as multivariate heteroscedastic nonlinear 
errors-in-variables models, may be attained via iterative formula~(\ref{Fisher-Scoring}).

It is well known that MLEs are consistent, asymptotically efficient and
asymptotically normal distributed. We can write
$\widehat{\bm{\theta}}\stackrel{a}{\sim}\mathcal{N}_{p}(\bm{\theta},
\bm{K}_{\bm{\theta}}^{-1})$, when $n$ is large, $\stackrel{a}{\sim}$ denoting approximately distributed.
Hence, hypotheses testing can be carried out using this asymptotic distribution.

\section{Biases of estimates of $\bm{\theta}$}

In this section we compute the biases of ML estimates of $\bm{\theta}$
for models defined by~(\ref{MainModel}).
Let $B(\widehat{\theta}_{a})$ be the $n^{-1}$ bias of $\widehat{\theta}_{a}$, $a=1,2,\ldots,p$.
It follows from the general expression for the multiparameter $n^{-1}$ biases of MLEs given
by \cite{CoxSnell1968} that
\[
B(\widehat{\theta}_{a}) = \sideset{}{^{\prime}}\sum_{t,s,r}\kappa^{a,t}\kappa^{s,r}\biggl\{
\frac{1}{2}\kappa_{tsr}-\kappa_{ts,r}\biggr\}, 
\]
where $\sum^{'}$ denotes the summation over all combinations of the parameters
$\theta_{1},\ldots, \theta_{p}$.
Following \cite{CordeiroKlein1994}, we write the above equation in matrix notation
to obtain $n^{-1}$ bias vector $\bm{B}(\widehat{\bm{\theta}})$ of $\widehat{\bm{\theta}}$ in the form
\[
\bm{B}(\widehat{\bm{\theta}}) = \bm{K}_{\bm{\theta}}^{-1}\bm{W}\vvec(\bm{K}_{\bm{\theta}}^{-1}), 
\]
where $\bm{W} = (\bm{W}^{(1)},\ldots,\bm{W}^{(p)})$ is a $p\times p^{2}$ partitioned matrix,
each $\bm{W}^{(r)}$ referring to the $r^{th}$ component of $\bm{\theta}$
being a $p\times p$ matrix with typical $(t,s)$th element given by
$w_{ts}^{(r)} = \frac{1}{2}\kappa_{tsr}+\kappa_{ts,r} = \kappa_{ts}^{(r)}
-\frac{1}{2}\kappa_{tsr}$. Notice that from 
(\ref{krst}) and (\ref{krs(t)}) we have that
\begin{align}\label{wts(r)}
\begin{split}
w_{ts}^{(r)} &=\frac{1}{4}\tr\{\bm{A}_{t}\bm{C}_{sr}  + \bm{A}_{s}\bm{C}_{tr} - \bm{A}_{r}\bm{C}_{ts}\}\\
&\quad-\frac{1}{2}(\bm{a}_{t}^{\top}\bm{\Sigma}^{-1}\bm{a}_{sr}+\bm{a}_{s}^{\top}\bm{\Sigma}^{-1}\bm{a}_{tr}
 - \bm{a}_{r}^{\top}\bm{\Sigma}^{-1}\bm{a}_{ts})\\
&\quad+\frac{1}{2}(\bm{a}_{s}^{\top}\bm{A}_{t}\bm{a}_{r} + \bm{a}_{t}^{\top} \bm{A}_{s}\bm{a}_{r}
-\bm{a}_{t}^{\top}\bm{A}_{r}\bm{a}_{s}).
\end{split}
\end{align}
Since $\bm{K}_{\bm{\theta}}$ is a symmetric  matrix and we are interested in the multiplication
result of $\bm{W}\vvec(\bm{K}_{\bm{\theta}}^{-1})$, many terms of~(\ref{wts(r)}) cancel. Indeed, 
note that the $t^{th}$ element of $\bm{W}\vvec(\bm{K}_{\bm{\theta}}^{-1})$ is given by 
$w_{t1}^{(1)}\kappa^{1,1} + (w_{t2}^{(1)} + w_{t1}^{(2)})\kappa^{1,2} + \cdots +
(w_{tr}^{(s)} + w_{ts}^{(r)})\kappa^{s,r} +
\cdots   +(w_{tp}^{(p-1)} + w_{t(p-1)}^{(p)})\kappa^{p-1,p} +w_{tp}^{(p)}\kappa^{p,p}$ and 
$w_{tr}^{(s)} + w_{ts}^{(r)} = \frac{1}{2}\tr(\bm{A}_{t}\bm{C}_{sr})
- \bm{a}_{t}^{\top}\bm{\Sigma}^{-1}\bm{a}_{sr} +  \bm{a}_{s}^{\top}\bm{A}_{t}\bm{a}_{r}$.
Therefore, we can replace the element
$w_{ts}^{(r)}$ by $\frac{1}{4}\tr(\bm{A}_{t}\bm{C}_{sr})- \frac{1}{2}\bm{a}_{t}^{\top}\bm{\Sigma}^{-1}\bm{a}_{sr} + \frac{1}{2} \bm{a}_{s}^{\top}\bm{A}_{t}\bm{a}_{r}$
and $\bm{W}^{(r)}$ may be written in an equivalent way as
$\bm{W}^{(r)} = \widetilde{\bm{F}}^{\top}\widetilde{\bm{H}}\bm{\Phi}_{r}$, $r=1,\ldots, p$, where
$\bm{\Phi}_{r} = -\frac{1}{2}(\bm{G}_{r} + \bm{J}_r)$ with
\[
\bm{G}_{r}=
\begin{bmatrix}
\bm{a}_{1r} & \cdots & \bm{a}_{pr}\\
\vvec(\bm{C}_{1r})& \cdots &    \vvec(\bm{C}_{pr})
\end{bmatrix}\quad{\rm and}\quad
\bm{J}_r =
\begin{bmatrix}
\bm{0}    \\
2(\bm{I}_{nq}\otimes\bm{a}_r)\widetilde{\bm{D}}      
\end{bmatrix},
\]
where $\bm{I}_{m}$ denotes an $m\times m$ identity matrix.
That is, the matrix $\bm{W}$ can be written as
$\bm{W} = \widetilde{\bm{F}}^{\top}\widetilde{\bm{H}} (\bm{\Phi}_1,  \ldots,  \bm{\Phi}_p)$.
Then, we arrive at the following theorem.
\begin{teorema} The $n^{-1}$ bias vector $\bm{B}(\widehat{\bm{\theta}})$
of $\widehat{\bm{\theta}}$ is given by
\begin{equation}\label{BIAS-vector}
\bm{B}(\widehat{\bm{\theta}}) = (\widetilde{\bm{F}}^{\top}\widetilde{\bm{H}}\widetilde{\bm{F}})^{-1}
\widetilde{\bm{F}}^{\top}\widetilde{\bm{H}}\widetilde{\bm{\xi}},
\end{equation}
where $\widetilde{\bm{\xi}} = (\bm{\Phi}_1,\ldots,\bm{\Phi}_p)\vvec((\widetilde{\bm{F}}^{\top}
\widetilde{\bm{H}}\widetilde{\bm{F}})^{-1})$.
\end{teorema}

In order to interpret formulae~(\ref{BIAS-vector}) it is helpful
to exploit the relationship between the $n^{-1}$ bias of $\widehat{\bm{\theta}}$ and a linear regression.
The bias vector $\bm{B}(\widehat{\bm{\theta}})$ is simply the set coefficients from the ordinary weighted
lest-squares regression of $\widetilde{\bm{\xi}}$ on the columns of $\widetilde{\bm{F}}$,
using weights in $\widetilde{\bm{H}}$. As expression~(\ref{BIAS-vector})
makes clear, for any particular model of the class of models presented in Section~\ref{model-specification},
it is always possible to express the bias of $\widehat{\bm{\theta}}$ as the solution of an ordinary weighted
lest-squares regression.
Equation~(\ref{BIAS-vector}) is easily handled algebraically for any type
of nonlinear model, since it only involves simple operations on matrices and vectors. This
equation, in conjunction with a computer algebra system such as {\tt MAPLE} \citep{AbellBrase94},
will yield $\bm{B}(\widehat{\bm{\theta}})$ algebraically with minimal effort. Also, we
can compute the bias $\bm{B}(\widehat{\bm{\theta}})$ numerically via
software with numerical linear algebra facilities such as {\tt Ox} \citep{DcK2006} and
{\tt R} \citep{R2006}. [Note that we have described a procedure to attain a corrected estimator
in a general formulation that covers a wide class of models.
In the next section we shall present some special cases to shed light
on the applicability of our general formulation.]

Therefore, we are able to compute the $n^{-1}$ biases of the MLEs for the general model (\ref{MainModel})
using formula~(\ref{BIAS-vector}). On the right-hand side of expression~(\ref{BIAS-vector}), which is of order
$n^{-1}$, consistent estimates of the parameter $\bm{\theta}$ can be inserted to define
the corrected MLE $\widetilde{\bm{\theta}} = \widehat{\bm{\theta}} - \widehat{\bm{B}}(\widehat{\bm{\theta}})$,
where $\widehat{\bm{B}}(\cdot)$ denotes the value of $\bm{B}(\cdot)$ at $\widehat{\bm{\theta}}$.
The bias-corrected estimate $\widetilde{\bm{\theta}}$  is expected to have
better sampling properties than the uncorrected estimator, $\widehat{\bm{\theta}}$.
In fact, we present some simulations in Section~\ref{simulation} that indicate that
$\widetilde{\bm{\theta}}$ has smaller bias than its corresponding MLE,
thus suggesting that the bias corrections above have the effect of shrinking the modified estimates
toward to the true parameter values.

Following \cite{Cordeiro2008}, it is worth emphasizing that there are other methods to bias-correcting MLEs.
In regular parametric problems, \cite{Firth1993} developed the so-called ``preventive'' method,
which also allows for the removal of the second-order biases. His method consists of modifying the original score
function to remove the first-order term from the asymptotic bias of these estimates.
In exponential families with canonical parameterization his correction scheme
consists in penalizing the likelihood by the Jeffreys invariant prior.
This is a preventive approach to bias adjustment which has its merits, but the connections between our
results and his work are not pursued in this paper since they will be developed in future research.
Additionally, we should also stress that it is possible to avoid cumbersome and tedious
algebra on cumulant calculations by using Efron's bootstrap \citep{EfronTibs}.
We use the analytical approach here since it leads to a closed-form solution and we do not need to run 
extensive numerical resamples. Moreover, the application
of the analytical bias seems to generally be the most feasible procedure
to use and it continues to receive attention in the literature.

\section{Special models}

It is useful to consider some examples to illustrate
the applicability of the results in the previous section and clarify the
notation used. Other important special models could also be easily handled since formula~(\ref{BIAS-vector})
only requires simple operations on matrices and vectors.

First, consider a univariate nonlinear model ($q=1$) in which $\bm{\Sigma} = \sigma^2\bm{I}_{n}$.
Note that this model is a particular case of model~(\ref{MainModel})
with $\bm{\theta} = (\bm{\beta}^{\top}, \sigma^{2})^{\top}$ and
$\bm{\mu}=(\mu_1(\bm{\beta}), \ldots, \mu_n(\bm{\beta}))^{\top}$, where 
the components of $\bm{\mu}$ and $\bm{\Sigma}$
are unrelated and vary independently. Let $p-1$ be the dimension of $\bm{\beta}$.
Here, $\widetilde{\bm{D}} = (\bm{a}_1, \ldots, \bm{a}_{p-1}, \bm{0})$
and $\widetilde{\bm{V}} = (\bm{0},\vvec(\bm{C}_{p}))$.
Also, $\widetilde{\bm{F}} = {\rm diag}\{\widetilde{\bm{D}}^{(1)}, \widetilde{\bm{V}}^{(2)}\}$,
where $\widetilde{\bm{D}}^{(1)} = (\bm{a}_1, \ldots, \bm{a}_{p-1})$ and
$\widetilde{\bm{V}}^{(2)} = \vvec(\bm{C}_p)$.
Then, from~(\ref{InfFisher}), the expected Fisher information for $\bm{\theta}$ can be written as
$\bm{K}_{\bm{\theta}} = \widetilde{\bm{F}}^{\top}\widetilde{\bm{H}}\widetilde{\bm{F}} =
{\rm diag}\{\bm{K}_{\bm{\beta}}, K_{\sigma^2}\}$, where
$\bm{K}_{\bm{\beta}} = \widetilde{\bm{D}}^{(1)\top}\widetilde{\bm{D}}^{(1)}/\sigma^2$
is Fisher's information for $\bm{\beta}$ and
$K_{\sigma^2} = n/2\sigma^4$
is the information relative to $\sigma^2$.
Since $\bm{K}_{\bm{\theta}}$ is block-diagonal,
$\bm{\beta}$ and $\sigma$ are globally orthogonal \citep{CoxReid87}.
From~(\ref{BIAS-vector}), note that
\[
(\widetilde{\bm{F}}^{\top}\widetilde{\bm{H}}\widetilde{\bm{F}})^{-1}
\widetilde{\bm{F}}^{\top}\widetilde{\bm{H}} =
\begin{bmatrix}
(\widetilde{\bm{D}}^{(1)\top}\widetilde{\bm{D}}^{(1)})^{-1}\widetilde{\bm{D}}^{(1)\top} & \bm{0}\\
 \bm{0} & \frac{1}{n}\widetilde{\bm{V}}^{(2)\top}(\bm{I}_{n}\otimes\bm{I}_{n})
\end{bmatrix}.
\]
Also,
\[
\widetilde{\bm{\xi}} =
\begin{pmatrix}
\widetilde{\bm{\xi}}_{1} \\
\widetilde{\bm{\xi}}_{2}
\end{pmatrix}
=
\begin{bmatrix}
-\frac{\sigma^2}{2}\ddot{\bm{G}}\vvec\{(\widetilde{\bm{D}}^{(1)\top}\widetilde{\bm{D}}^{(1)})^{-1}\}\\
-\sum_{k=1}^{p-1}(\bm{I}_{n}\otimes\bm{a}_k)\widetilde{\bm{D}}^{(1)}\bm{K}_{\bm{\beta}k}^{-1}
\end{bmatrix},
\]
where $\ddot{\bm{G}} = (\bm{a}_{\bm{\beta}1},\ldots,\bm{a}_{\bm{\beta}(p-1)})$
with $\bm{a}_{\bm{\beta}k} = (\bm{a}_{1k}, \ldots,\bm{a}_{(p-1)k})$
and $\bm{K}_{\bm{\beta}k}^{-1}$ is the $k^{th}$ column of $\bm{K}_{\bm{\beta}}^{-1}$.
Then,
\[
\bm{B}(\widehat{\bm{\theta}}) =
\begin{pmatrix}
\bm{B}(\widehat{\bm{\beta}})\\
B(\widehat{\sigma}^2)\\
\end{pmatrix}=
\begin{bmatrix}
(\widetilde{\bm{D}}^{(1)\top}\widetilde{\bm{D}}^{(1)})^{-1}\widetilde{\bm{D}}^{(1)\top}\widetilde{\bm{\xi}}_{1}\\
\frac{1}{n}\widetilde{\bm{V}}^{(2)\top}(\bm{I}_{n}\otimes\bm{I}_{n})\widetilde{\bm{\xi}}_{2}
\end{bmatrix}.
\]
Note that $\bm{B}(\widehat{\bm{\beta}}) = (\widetilde{\bm{D}}^{(1)\top}
\widetilde{\bm{D}}^{(1)})^{-1}\widetilde{\bm{D}}^{(1)\top}\widetilde{\bm{\xi}}_{1}$
agrees with the result due to Cook et al.~(1986, Equation~(3)). Additionally,
we obtain the following simple form originally first given by \cite{Beale60}:
$B(\widehat{\sigma}^2) = -(p-1)\sigma^2/n$;
note that
\begin{align*}
\widetilde{\bm{V}}^{(2)\top}(\bm{I}_{n}\otimes\bm{I}_{n})
\sum_{k=1}^{p-1}(\bm{I}_{n}\otimes\bm{a}_k)\widetilde{\bm{D}}^{(1)}\bm{K}_{\bm{\beta}k}^{-1} &=
\sum_{k=1}^{p-1}\vvec(\bm{C}_{p})^{\top}(\bm{I}_{n}\otimes\bm{a}_k)\widetilde{\bm{D}}^{(1)}\bm{K}_{\bm{\beta}k}^{-1}\\
&\hspace{-4.3cm}= \sum_{k=1}^{p-1}\tr\{\bm{a}_k\bm{K}_{\bm{\beta}k}^{-1}\widetilde{\bm{D}}^{(1)\top}\}
= \tr\{\widetilde{\bm{D}}^{(1)}\bm{K}_{\bm{\beta}}^{-1}\widetilde{\bm{D}}^{(1)\top}\} = (p-1)\sigma^2.
\end{align*}

As a second application, consider the multivariate nonlinear heteroscedastic
regression model studied by \cite{Vasccor1997}. 
Note that this model is a particular case of model~(\ref{MainModel}),
with $\bm{\theta} = (\bm{\beta}^{\top}, \bm{\sigma}^{\top})^{\top}$,
$\bm{\mu}=\vvec(\bm{\mu}_1(\bm{\beta}), \ldots, \bm{\mu}_n(\bm{\beta}))$ and 
$\bm{\Sigma}=\mbox{diag}\{\bm{\Sigma}_1(\bm{\sigma}), \ldots,\bm{\Sigma}_n(\bm{\sigma})\} $. Therefore, 
the components of $\bm{\mu}$ and $\bm{\Sigma}$
are unrelated and vary independently. Let $p_1$ and $p_2 = p - p_{1}$ be the dimensions of $\bm{\beta}$
and $\bm{\sigma}$, respectively.
Here, $\widetilde{\bm{D}} = (\bm{a}_1, \ldots, \bm{a}_{p_{1}}, \bm{0})$
and $\widetilde{\bm{V}} = (\bm{0},\vvec(\bm{C}_{p_{1}+1}),\ldots, \vvec(\bm{C}_p))$.
Let $\widetilde{\bm{D}}^{(1)} = (\bm{a}_1,\bm{a}_2, \ldots, \bm{a}_{p_{1}})$ and
$\widetilde{\bm{V}}^{(2)} = (\vvec(\bm{C}_{p_{1}+1}),\vvec(\bm{C}_{p_{1}+2})\ldots, \vvec(\bm{C}_p))$, then
$\widetilde{\bm{F}} = {\rm diag}\{\widetilde{\bm{D}}^{(1)}, \widetilde{\bm{V}}^{(2)}\}$.
From~(\ref{InfFisher}), the expected Fisher information for $\bm{\theta}$ can be written as
$\bm{K}_{\bm{\theta}} = \widetilde{\bm{F}}^{\top}\widetilde{\bm{H}}\widetilde{\bm{F}} =
{\rm diag}\{\bm{K}_{\bm{\beta}}, \bm{K}_{\bm{\sigma}}\}$, where
$\bm{K}_{\bm{\beta}} = \widetilde{\bm{D}}^{(1)\top}\bm{\Sigma}^{-1}\widetilde{\bm{D}}^{(1)}$
is Fisher's information for $\bm{\beta}$ and
$\bm{K}_{\bm{\sigma}} = \frac{1}{2}\widetilde{\bm{V}}^{(2)\top}\widetilde{\bm{\Sigma}}^{-1}\widetilde{\bm{V}}^{(2)}$
is the information relative to $\bm{\sigma}$.
Since $\bm{K}_{\bm{\theta}}$ is block-diagonal,
$\bm{\beta}$ and $\bm{\sigma}$ are globally orthogonal.
From~(\ref{BIAS-vector}), it follows that
\[
(\widetilde{\bm{F}}^{\top}\widetilde{\bm{H}}\widetilde{\bm{F}})^{-1}
\widetilde{\bm{F}}^{\top}\widetilde{\bm{H}} =
\begin{bmatrix}
(\widetilde{\bm{D}}^{(1)\top}\bm{\Sigma}^{-1}\widetilde{\bm{D}}^{(1)})^{-1}\widetilde{\bm{D}}^{(1)\top}\bm{\Sigma}^{-1} & \bm{0}\\
 \bm{0} & (\widetilde{\bm{V}}^{(2)\top}\widetilde{\bm{\Sigma}}^{-1}\widetilde{\bm{V}}^{(2)})^{-1}\widetilde{\bm{V}}^{(2)\top}\widetilde{\bm{\Sigma}}^{-1}
\end{bmatrix}.
\]
Also,
\[
\widetilde{\bm{\xi}} =
\begin{pmatrix}
\widetilde{\bm{\xi}}_{1} \\
\widetilde{\bm{\xi}}_{2}
\end{pmatrix}
=
\begin{bmatrix}
-\frac{1}{2}\ddot{\bm{G}}\vvec\{(\widetilde{\bm{D}}^{(1)\top}\bm{\Sigma}^{-1}\widetilde{\bm{D}}^{(1)})^{-1}\}\\
-\bigl(\ddot{\bm{W}}\vvec\{(\widetilde{\bm{V}}^{(2)\top}\widetilde{\bm{\Sigma}}^{-1}\widetilde{\bm{V}}^{(2)})^{-1}\}
+ \sum_{k=1}^{p_{1}}(\bm{I}_{nq}\otimes\bm{a}_k)\widetilde{\bm{D}}^{(1)}\bm{K}_{\bm{\beta}k}^{-1}\bigr)
\end{bmatrix},
\]
where $\ddot{\bm{G}} = (\bm{a}_{\bm{\beta}1},\ldots,\bm{a}_{\bm{\beta}p_{1}})$
with $\bm{a}_{\bm{\beta}k} = (\bm{a}_{1k}, \ldots,\bm{a}_{p_{1}k})$ and
$\ddot{\bm{W}} = (\bm{v}_{\bm{\sigma}(p_{1}+1)},\ldots,\bm{v}_{\bm{\sigma}p})$ with
$\bm{v}_{\bm{\sigma}k} = (\vvec(\bm{C}_{(p_{1}+1)k}),\ldots, \vvec(\bm{C}_{pk}))$
and $\bm{K}_{\bm{\beta}k}^{-1}$ is the $k^{th}$ column of $\bm{K}_{\bm{\beta}}^{-1}$.
Therefore,
\[
\bm{B}(\widehat{\bm{\theta}}) =
\begin{pmatrix}
\bm{B}(\widehat{\bm{\beta}})\\
\bm{B}(\widehat{\bm{\sigma}})\\
\end{pmatrix}=
\begin{bmatrix}
(\widetilde{\bm{D}}^{(1)\top}\bm{\Sigma}^{-1}\widetilde{\bm{D}}^{(1)})^{-1}\widetilde{\bm{D}}^{(1)\top}\bm{\Sigma}^{-1}\widetilde{\bm{\xi}}_{1}\\
(\widetilde{\bm{V}}^{(2)\top}\widetilde{\bm{\Sigma}}^{-1}\widetilde{\bm{V}}^{(2)})^{-1}\widetilde{\bm{V}}^{(2)\top}\widetilde{\bm{\Sigma}}^{-1}\widetilde{\bm{\xi}}_{2}
\end{bmatrix}.
\]
Note that $\bm{B}(\widehat{\bm{\beta}}) = (\widetilde{\bm{D}}^{(1)\top}\bm{\Sigma}^{-1}
\widetilde{\bm{D}}^{(1)})^{-1}\widetilde{\bm{D}}^{(1)\top}\bm{\Sigma}^{-1}\widetilde{\bm{\xi}}_{1}$
agrees with the result due to Vasconcellos and Cordeiro (1997, Equation~(3.2)). Additionally,
note that
$\bm{B}(\widehat{\bm{\sigma}}) = (\widetilde{\bm{V}}^{(2)\top}\widetilde{\bm{\Sigma}}^{-1}\widetilde{\bm{V}}^{(2)})^{-1}\widetilde{\bm{V}}^{(2)\top}
\widetilde{\bm{\Sigma}}^{-1}\widetilde{\bm{\xi}}_{2}$ also
reduces to Vasconcellos and Cordeiro's (1997) Eq.~(3.8), since
\[
\widetilde{\bm{V}}^{(2)\top}\widetilde{\bm{\Sigma}}^{-1}\sum_{k=1}^{p_{1}}(\bm{I}_{nq}\otimes\bm{a}_k)\widetilde{\bm{D}}^{(1)}\bm{K}_{\bm{\beta}k}^{-1} =
\widetilde{\bm{V}}^{(2)\top}\widetilde{\bm{\Sigma}}^{-1}\vvec(\bm{\Delta}_{*}),
\]
where $\bm{\Delta}_{*}$ is as defined by Vasconcellos and Cordeiro (1997, p.~148).

Next, unlike the two models discussed previously,
we consider a model where the elements of $\bm{\mu}$ and $\bm{\Sigma}$
are related and do not vary independently.
Consider the nonlinear heteroscedastic errors-in-variables model
\[
Y_{i} = \alpha + \beta x_{i} + \exp(\gamma z_{i}) + e_{i}\quad{\rm and}\quad
X_{i} = x_{i} + u_{i},
\]
where $x_i \sim \mathcal{N}(\mu_x, \sigma_x^2)$ and $u_i \sim\mathcal{N}(0,\sigma^2_u)$
are the measurement errors with $\sigma_u^2$
known and $e_i \sim \mathcal{N}(0, \sigma^2 \exp\{\eta z_i \})$.
The covariate $z_{i}$ is known. In this example, the vector of parameters is 
$\bm{\theta} = (\alpha, \beta, \gamma, \mu_x, \sigma_x^2, \sigma^2,\eta)^{\top}$
and the mean and variance functions for the $i^{th}$ observation ($Y_i,X_i$) are given by
\[
\bm{\mu}_i =
\begin{pmatrix}
\alpha + \beta \mu_x + \exp(\gamma z_i)\\
\mu_x
\end{pmatrix} 
\quad {\rm and} \quad
\bm{\Sigma}_i = 
\begin{pmatrix}
\beta^2 \sigma_x^2 +\sigma^2\exp(\eta z_i)& \beta \sigma_x^2\\
\beta \sigma_x^2   & \sigma_x^2 + \sigma_u^2
\end{pmatrix}.
\]
Then,
\[
\bm{a}_1 =  \bm{1}_n \otimes
\begin{pmatrix}
1\\
0
\end{pmatrix}, \quad
\bm{a}_2 =  \bm{1}_n \otimes
\begin{pmatrix}
\mu_x\\
0
\end{pmatrix}, \quad
\bm{a}_3 = \mbox{vec}
\left\{
\begin{pmatrix}
z_1 \exp(\gamma z_1)\\
0
\end{pmatrix} \cdots 
\begin{pmatrix}
z_n \exp(\gamma z_n)\\
0
\end{pmatrix}\right\},
\]
\[
\bm{a}_4 =  \bm{1}_n \otimes
\begin{pmatrix}
\beta\\
1
\end{pmatrix} \quad \mbox{and}\quad \bm{a}_{5}=\bm{a}_6=\bm{a}_7 = \bm{0},
\]
where $\bm{1}_{n}$ denotes an $n\times 1$ vector of ones.
Also, $\bm{a}_{rs} = \bm{0}$ for all $r, s$ except for
\[
\bm{a}_{24} = \bm{a}_{42} =  \bm{1}_n \otimes
\begin{pmatrix}
1\\
0
\end{pmatrix} \quad \mbox{and} \quad 
\bm{a}_{33} = \mbox{vec}
\left\{
\begin{pmatrix}
z_1^2 \exp(\gamma z_1)\\
0
\end{pmatrix} \cdots
\begin{pmatrix}
z_n^2 \exp(\gamma z_n)\\
0
\end{pmatrix}\right\}.
\]
Also, $\bm{C}_{r} = \bm{0}$ for all $r$ except for
\[
\bm{C}_{2} = \bm{I}_n \otimes
\begin{pmatrix}
2\beta\sigma_x^2 & \sigma_x^2\\
\sigma_x^2  &0
\end{pmatrix}, \
\bm{C}_{5} = \bm{I}_n \otimes
\begin{pmatrix}
\beta^2 & \beta\\
\beta  & 1
\end{pmatrix},\quad
\bm{C}_{6} = \sumdir
\begin{pmatrix}
\exp(\eta z_i) & 0\\
0  & 0
\end{pmatrix}
\]
and
\[ 
\bm{C}_{7} = \sumdir
\begin{pmatrix}
z_i\sigma^2\exp(\eta z_i) & 0\\
0  & 0
\end{pmatrix},
\]
where $\oplus$ is the direct sum of matrices. 
Additionally, $\bm{C}_{rs} = \bm{0}$ for all $r, s$  except for
\[
\bm{C}_{22} = \bm{I}_n \otimes
\begin{pmatrix}
2\sigma_x^2 & 0\\
0  & 0
\end{pmatrix},\quad
\bm{C}_{25} = \bm{C}_{52} = \bm{I}_n \otimes
\begin{pmatrix}
2\beta & 1\\
1  & 0
\end{pmatrix},
\]
\[
\bm{C}_{67} = \bm{C}_{76} =
\sumdir
\begin{pmatrix}
z_i\exp(\eta z_i) & 0\\
0  & 0
\end{pmatrix} \quad \mbox{and} \quad 
\bm{C}_{77} =
\sumdir
\begin{pmatrix}
z_i^2\sigma^2\exp(\eta z_i) & 0\\
0  & 0
\end{pmatrix}. 
\]
Thus, $\widetilde{\bm{D}} = (\bm{a}_1, \bm{a}_2, \bm{a}_3,\bm{a}_4,\bm{0},\bm{0},\bm{0})$,
$\widetilde{\bm{V}} = (\bm{0}, \vvec(\bm{C}_2), \bm{0},\bm{0},\vvec(\bm{C}_5),\vvec(\bm{C}_6),\vvec(\bm{C}_7))$ and
the matrix formula~(\ref{BIAS-vector}) can be used to compute the second-order bias
for this model. Notice that, as $\vvec(\bm{C}_2)$ is not equal to zero, the derivation of
algebraic expression using matrix formula~(\ref{BIAS-vector}) becomes very tedious, since
the structure of $\bm{K}_{\bm{\theta}}$ is not block-diagonal unlike the two  previous examples.
However, using {\tt MAPLE}, for example, the derivation can be easily done.
Also, the $n^{-1}$ bias vector $\bm{B}(\widehat{\bm{\theta}})$ can be obtained
numerically via software with numerical linear algebra facilities with minimal effort
such as {\tt R} and {\tt Ox}.

\section{Simulation study}\label{simulation}

We recall that, for large samples the biases of the MLEs are neglible. However, for small 
and moderate sample sizes the second-order biases may be large and can be used to improve
the estimation. We shall use Monte Carlo simulation to evaluate the finite sample performance
of the original MLEs and their corrected versions.
All simulations were performed using {\tt R} (R Development Core Team, 2006).
The sample sizes considered were $n= 15, 25, 35, 50$ and 100, 
the number of Monte Carlo replications was 5,000.

We consider the simple errors-in-variables model as described in \cite{Fuller}:
\[
Y_{i} = \alpha + \beta x_{i} + e_{i}\quad{\rm and}\quad
X_{i} = x_{i} + u_{i},
\]
where $x_i \sim \mathcal{N}(\mu_x, \sigma_x^2)$ and $u_i \sim\mathcal{N}(0,\sigma^2_u)$
are the measurement errors with $\sigma_u^2$
known and $e_i \sim \mathcal{N}(0, \sigma^{2})$, with $i=1,2,\ldots,n$.
Here, $\bm{\theta} = (\alpha, \beta,  \mu_x, \sigma_x^2, \sigma^{2})^{\top}$ and
\[
\bm{\mu} = \bm{1}_n \otimes 
\begin{pmatrix}
\alpha + \beta \mu_x\\
\mu_x
\end{pmatrix}
\quad {\rm and} \quad 
\bm{\Sigma} = \bm{I}_n \otimes 
\begin{pmatrix}
\beta^2 \sigma_x^2 +\sigma^2& \beta \sigma_x^2\\
\beta \sigma_x^2   & \sigma_x^2 + \sigma_u^2
\end{pmatrix}.
\]
From the previous expressions, we have immediately that
\[
\bm{a}_1 =  \bm{1}_n \otimes 
\begin{pmatrix}
1\\
0
\end{pmatrix}, \quad
\bm{a}_2 =  \bm{1}_n \otimes 
\begin{pmatrix}
\mu_x\\
0
\end{pmatrix}, \quad
\bm{a}_3 =  \bm{1}_n \otimes 
\begin{pmatrix}
\beta\\
1
\end{pmatrix}, \quad
\bm{a}_4 = \bm{a}_5 = \bm{0}
\]
and $\bm{a}_{rs} = \bm{0}$ for all $r, s$ except for
\[
\bm{a}_{23} = \bm{a}_{32} =  \bm{1}_n \otimes 
\begin{pmatrix}
1\\
0
\end{pmatrix}.
\]
Also, $\bm{C}_{r} = \bm{0}$ for all $r$ except for
\[
\bm{C}_{2} = \bm{I}_n \otimes 
\begin{pmatrix}
2\beta\sigma_x^2 & \sigma_x^2\\
\sigma_x^2  &0 
\end{pmatrix}, \
\bm{C}_{4} = \bm{I}_n \otimes 
\begin{pmatrix}
\beta^2 & \beta\\
\beta  & 1
\end{pmatrix}\quad\mbox{and}\quad
\bm{C}_{5} = \bm{I}_n \otimes 
\begin{pmatrix}
1 & 0\\
0  & 0
\end{pmatrix}.
\]
Additionally, $\bm{C}_{rs} = \bm{0}$ for all $r, s$  except for
\[
\bm{C}_{22} = \bm{I}_n \otimes 
\begin{pmatrix}
2\sigma_x^2 & 0\\
0  & 0
\end{pmatrix} \ \mbox{and} \ 
\bm{C}_{24} = \bm{C}_{42} = \bm{I}_n \otimes 
\begin{pmatrix}
2\beta & 1\\
1  & 0
\end{pmatrix}.
\]
Thus, $\widetilde{\bm{D}} = (\bm{a}_1, \bm{a}_2, \bm{a}_3,\bm{0},\bm{0})$ and
$\widetilde{\bm{V}} = (\bm{0}, \vvec(\bm{C}_2), \bm{0},\vvec(\bm{C}_4),\vvec(\bm{C}_5))$.
Therefore, all the quantities necessary to calculate $\bm{B}(\widehat{\bm{\theta}})$ using
expression~(\ref{BIAS-vector}) are given.

In order to analyze the point estimation results, we computed,
for each sample size and for each estimator: the relative bias (the relative bias of an estimator
$\widehat{\theta}$ is defined as $\{\Es(\widehat{\theta}) - \theta\}/\theta$,                    
its estimate being obtained by estimating $\Es(\widehat{\theta})$ by Monte Carlo)                
and the root mean square error, i.e., $\sqrt{{\rm MSE}}$, where MSE is the mean squared error        
estimated from the 5,000 Monte Carlo replications.                                              
Without loss of generality, the true values of the regression parameters were set at               
$\alpha=67$, $\beta=0.42$, $\mu_x=70$, $\sigma_x^2=247$ and $\sigma^2=43$.
The parameter setting were choosen in order to represent the dataset
(yields of corn on Marshall soil in Iowa)
presented in Fuller (1987, p.~18). The known measurement error variance
is $\sigma_u^2 = 57$ (which was attained through a previous experiment).

Table~\ref{tab:1} gives the relative biases and $\sqrt{{\rm MSE}}$
of both uncorrected and corrected estimates.
The figures in this table confirm that the bias-corrected estimates are generally closer to the true
parameter values than the unadjusted estimates. We observe that, in absolute value, the estimated
relative biases of the bias-corrected estimator were smaller than that of the original MLE for
all sample sizes considered, thus showing the effectiveness of the bias
correction schemes used in the definition of these estimators.

For instance, when $n=15$, the estimated relative bias of the estimators of $\alpha$,
$\beta$, $\mu_x$, $\sigma_x^2$ and $\sigma^2$ 
average $-0.0518$ whereas the biases of the five corresponding
bias-adjusted estimators average $-0.0056$; that is, the average bias (in value absolute) of the MLEs is
almost ten times larger than that of the bias-corrected estimators.
This suggests that the second-order bias of MLEs should not be ignored in samples
of small to moderate sizes since they can be nonnegligible.

We can readily see that the MLEs of $\sigma_x^2$ and $\sigma^2$ are
on average far from the true parameter value, thus displaying large bias, for the different sample sizes
considered, even when $n=100$. This stresses the importance of using a bias correction.
For instance, when $n=50$, the relative biases of
$\widehat{\sigma}_{x}^{2}$ and $\widehat{\sigma}^{2}$ (MLEs)
were $-0.0226$ and $-0.0563$,  respectively, while the relative biases of
$\widetilde{\sigma}_{x}^{2}$ and $\widetilde{\sigma}^{2}$ (BCEs)
were $ 0.0016$ (sixteen times lesser) and $-0.0011$ (fifty times lesser), respectively. Observe that the MLEs
are always underestimating the true values of $\sigma_{x}^{2}$ and $\sigma^{2}$,
since their biases are always negatives. Note also that
root mean square error decrease with $n$, as expected.
Additionally, we note that all estimators have similar root mean squared errors.
\begin{table}[!htp]\renewcommand{\arraystretch}{1.16}
\begin{center}
\caption{Relative biases and $\sqrt{{\rm MSE}}$ of uncorrected and corrected 
estimates for an errors-in-variables model.}\label{tab:1}
\begin{tabular}{ccrrrrr}\hline
 &       & \multicolumn{2}{c}{MLE} & &\multicolumn{2}{c}{BCE}\\\cline{3-4}\cline{6-7}
$n$ &  $\bm{\theta}$ &Rel.~bias & $\sqrt{{\rm MSE}}$ && Rel.~bias & $\sqrt{{\rm MSE}}$  \\\hline 
15&  $\alpha$       &  $-0.0240$        &   12.46  &&   $ 0.0232$       &   11.29    \\                  
 &  $\beta$          &  $ 0.0547$        &    0.17  &&   $-0.0526$       &    0.16    \\                     
 &  $\mu_{x}$        &  $ 0.0014$        &    4.48  &&   $ 0.0014$       &    4.48    \\                     
 &  $\sigma_{x}^2$   &  $-0.0796$        &  108.49  &&   $-0.0029$       &  113.81    \\                     
 &  $\sigma^2$       &  $-0.1807$        &   19.52  &&   $ 0.0031$       &   20.38    \\\hline                                   
25 &  $\alpha$       &  $-0.0198$        &    9.05  &&   $ 0.0009$       &    8.14  \\                                                
 &  $\beta$          &  $ 0.0440$        &    0.13  &&   $-0.0029$       &    0.11  \\                                           
 &  $\mu_{x}$        &  $ 0.0004$        &    3.43  &&   $ 0.0004$       &    3.43  \\                            
 &  $\sigma_{x}^2$   &  $-0.0553$        &   85.73  &&   $-0.0082$       &   88.05  \\                      
 &  $\sigma^2$       &  $-0.1198$        &   15.48  &&   $-0.0104$       &   15.73  \\\hline                                         
35 &  $\alpha$       &  $-0.0117$        &    7.05  &&   $ 0.0010$       &    6.68  \\                                     
 &  $\beta$          &  $ 0.0267$        &    0.10  &&   $-0.0023$       &    0.09  \\                                   
 &  $\mu_{x}$        &  $-0.0001$        &    2.96  &&   $-0.0001$       &    2.96  \\                                   
 &  $\sigma_{x}^2$   &  $-0.0424$        &   71.36  &&   $-0.0084$       &   72.64  \\                             
 &  $\sigma^2$       &  $-0.0799$        &   12.83  &&   $-0.0014$       &   13.04  \\\hline                                                                  
50 &  $\alpha$       &  $-0.0080$        &    5.69  &&   $ 0.0002$       &    5.50  \\                                              
 &  $\beta$          &  $ 0.0190$        &    0.08  &&   $ 0.0005$       &    0.08  \\                                           
 &  $\mu_{x}$        &  $-0.0007$        &    2.45  &&   $-0.0007$       &    2.45  \\                                      
 &  $\sigma_{x}^2$   &  $-0.0226$        &   60.76  &&   $ 0.0016$       &   61.71  \\                             
 &  $\sigma^2$       &  $-0.0563$        &   10.75  &&   $-0.0011$       &   10.89  \\\hline
100 &  $\alpha$      &  $-0.0025$        &    3.83  &&   $ 0.0013$       &    3.78  \\                                              
 &  $\beta$          &  $ 0.0057$        &    0.05  &&   $-0.0029$       &    0.05  \\                                           
 &  $\mu_{x}$        &  $ 0.0002$        &    1.72  &&   $ 0.0002$       &    1.72  \\                                      
 &  $\sigma_{x}^2$   &  $-0.0131$        &   42.24  &&   $-0.0009$       &   42.54  \\                             
 &  $\sigma^2$       &  $-0.0298$        &    7.63  &&   $-0.0021$       &    7.67  \\\hline
\multicolumn{6}{l}{BCE: bias-corrected estimator.}                                              
\end{tabular}                                                                               
\end{center}                                                                                        
\end{table}

\section{An empirical ilustration}\label{application}

Next, as an empirical ilustration, consider a small data set given by Fuller (1987, p.~18).
The data set is presented in Table~\ref{dados}.
The data are yields of corn and determinations of available soil nitrogen collected at 11
sites on Marshall soil in Iowa. Following Fuller (1987, p.~18), we assume that
the estimates of soil nitrogen contain measurement
erros arise from two sources. First, only a small sample of soil is selected from each
plot and, as a result, there is the sampling error associated with the use of sample
to represent the whole. Second, there is a measurement error associated with the chemical
analysis used to determined the level of nitrogen in the soil sample. The variance
arising from these two sources is $\sigma_{u}^{2} = 57$.
According to Fuller (1987, p.~18), model~(\ref{EVM}) is a valid representation to these data.
\begin{table}[!htp]\renewcommand{\arraystretch}{0.94}
\begin{center}
\caption{Yields of corn on Marshall soil in Iowa.}\label{dados}
\begin{tabular}{ccc|ccc}\hline
       &         &  Soil     &         &         &  Soil     \\                     
       & Yield   &  Nitrogen &         & Yield   &  Nitrogen \\                     
Site   &  $(Y)$  &  $(X)$    &  Site   &  $(Y)$  &  $(X)$    \\\hline                     
1 & 86 & 70  & 7  &  99 & 50 \\                     
2 &115 & 97  & 8  &  96 & 70 \\                     
3 & 90 & 53  & 9  &  99 & 94 \\                     
4 & 86 & 64  & 10 & 104 & 69 \\                     
5 &110 & 95  & 11 &  96 & 51 \\                     
6 & 91 & 64  &   &   &  \\\hline                     
\end{tabular}                                                                 
\end{center}                                                                                        
\end{table}

The MLEs, the large-sample estimates of the corresponding standard errors, the
biases and the bias-corrected estimates are given Table~\ref{applic}.
These estimates were obtained using {\tt R}. 
The figures in this table show that the biases of the estimates of $\alpha$ and
$\beta$ are much less than standard errors of the corresponding estimates.
In cases of marginal statistical significance, biases of this maginitude could have a small
effect on the conclusions.  However, note that the MLEs of $\sigma_{x}^2$ and
$\sigma^2$ are strongly biased, as evidenced by our simulations studies,
i.e., they underestimate the model variances. Therefore, the bias-corrected estimates may be used
instead of the MLEs to make point inferences. 
\begin{table}[!htp]\renewcommand{\arraystretch}{1.1}
\begin{center}
\caption{MLEs and bias-corrected estimatives.}\label{applic}
\begin{tabular}{crrrr}\hline
Parameter       &  MLEs  &  S.E.   &   Bias    &   BCEs        \\\hline
$\alpha$        & 66.8606& 11.7272 & $  -2.5334$&    69.3939 \\
$\beta$         &  0.4331&  0.1633 & $   0.0359$&     0.3973 \\
$\mu_{x}$       & 70.6364&  5.0194 & $   0.0000$&    70.6364 \\
$\sigma_{x}^2$  &220.1405&118.1731 & $ -25.1946$&   245.3351 \\
$\sigma^2$      & 38.4058& 20.9357 & $ -10.3344$&    48.7402 \\\hline
\multicolumn{5}{l}{BCE: bias-corrected estimate.}                                              
\end{tabular}                                                                 
\end{center}                                                                                        
\end{table}

\begin{figure}
\centering
\includegraphics[width=11.6cm, height=8.6cm]{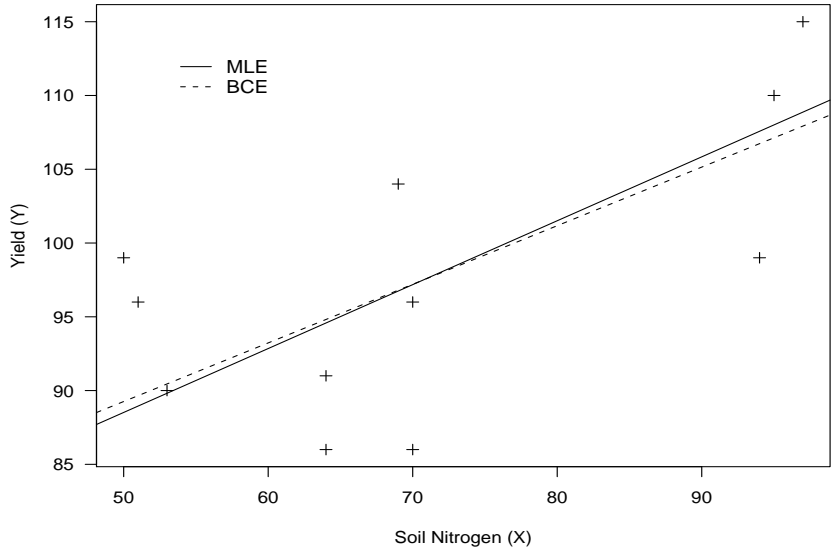}
\caption{Scatterplot for the data set together with the fitted lines.}\label{fig}
\end{figure}

Figure~\ref{fig} presents the scatterplot of the data
together with the fitted lines obtained using the MLEs and BCEs. Notice
that the line produced by the bias correction scheme the inclination slightly attenuated
and intercept increased relative to the non-corrected one.

\section{Conclusions}\label{conclusion}                                                  
                                                                                   
This paper proposed a bias correction for a multivariate normal model           
with a quite general parameterization. The main result centers on models where          
the mean and the variance share the same vector of parameters. Many models         
are particular cases of the proposed model such as (non)linear regressions, errors-in-variables 
models, mixed models, factor analysis and so forth. We have shown that
it is always possible to express the second order bias vector of the maximum likelihood estimates
as an ordinary weighted least-squares regression. Moreover, we derived a bias-adjustment
scheme that nearly eliminates the bias of the maximum likelihood estimator in small
and moderate samples. Our simulation results suggest that the bias correction we have
derived is very effective, even when the sample size is small.
Indeed, the bias correction mechanism
proposed in this paper yields modified maximum likelihood estimators that are
nearly unbiased. We also presented an
empirical ilustration that illustrates the proposed bias-adjustment scheme.

\section*{Acknowledgments}

We gratefully acknowledge grants from FAPESP. We wish to thank Francisco Cribari-Neto for valuable comments on this manuscript.

\end{document}